# Structural phenomena associated with the spin-state transition in LaCoO$_3$

P. G. Radaelli[a], and S-W. Cheong[b]

[a] ISIS facility, Rutherford Appleton Laboratory-CCLRC, Chilton, Didcot, Oxfordshire, OX11 0QX, United Kingdom

[b] Department of Physics and Astronomy, Rutgers University, Piscataway, NJ 08855 USA

ABSTRACT

The structural properties of LaCoO$_3$ were studied by means of high-resolution neutron powder diffraction in the temperature range $5 \leq T \leq 1000$ K. Changes of the Co$^{+3}$ spin states in this temperature interval are shown to affect not only the unit cell volume, as previously known, but also internal parameters such as the metal-oxygen bond lengths. These data, as well as the temperature-dependent magnetic susceptibility, can be qualitatively modeled based on a 3-state (low-spin, intermediate-spin and high-spin) activated behavior, but correction terms are required for quantitative agreement. Our fits consistently indicate that the ionic radius of the intermediate-spin state (~0.56 Å) is smaller than the low-spin/high-spin average (~0.58 Å). We also present evidence of a *third* lattice, anomaly, occurring around 800 K, which we attribute to the formation of oxygen vacancies.

1. INTRODUCTION

The lattice, electronic and magnetic properties of LaCoO$_3$ have been the subject of continuing interest since the fifties [1-7]. The core of the debate is the interpretation of two broad transitions in the magnetic susceptibility, occurring near 100 and 500 K. The effective bulk magnetic moment, which is zero at low temperatures once the effect of magnetic impurities and the surface is subtracted, increases rapidly to about 2.5 $\mu_B$ at the first transition, followed by a broad plateau. A second increase of the magnetic susceptibility to ~3.5 $\mu_B$ occurs at the second transition ($T_{II}$ ~ 500 K), which is associated with a metal-insulator transition. Both transitions are associated with anomalies in the structural parameters[8] and in the thermopower[9]. The classic interpretation of these effects, first given by Goodenough [2], invokes the existence of two Co spin states: $Co^{+3}_{LS}$ ($t^6_{2g}$, S=0) and $Co^{+3}_{HS}$ ($t^4_{2g}e^2_g$, S=2). At the first transition, the thermal population of $Co^{+3}_{HS}$ would increase up to a value of approximately 50 % at the expense of $Co^{+3}_{LS}$. A further increase of the population of $Co^{+3}_{HS}$ would be inhibited at first by the formation of an ordered superstructure of $Co^{+3}_{LS}$ and $Co^{+3}_{HS}$. In this framework, the second high-temperature anomaly is interpreted as an order-disorder transition where the $e_g$ electrons would gradually become mobile and destroy the superstructure.

The presence of an ordered phase in the intermediate temperature range is a crucial ingredient of this model. However, detailed neutron diffraction studies[10, 11] have failed to detect the expected symmetry change from $R\bar{3}c$ to $R\bar{3}$, associated with the ordered phase. Therefore, the theory was modified [9] to exclude long-range-ordering, in favor of a phase in which dynamic $Co^{+3}_{LS}/Co^{+3}_{HS}$ ordering would occur with a short correlation radius.

A radically different interpretation emerged from experimental studies of the electronic structure of LaCoO$_3$, such as electron photo-emission spectroscopy[12] and x-ray absorption spectroscopy[13, 14]. These data have been difficult to reconcile with the magnetization results, because significant changes in the spectra are only observed at the high-temperature transition. In a 1995 study of the related compound SrCoO$_3$, Potze and coworkers proposed that an *intermediate* spin (IS) state ($t^5_{2g}e^1_g$, S=2) could be favored by the relative stability of the ligand hole state, both for the Co$^{+4}$ and the Co$^{+3}$ species[15]. This proposal led to a thorough re-examination of the spectroscopic, magnetic and structural data (see for example [14]). Based on the most recent results, a consensus is building around the concept that all three spin states, LS, IS, and HS, are relevant for LaCoO$_3$. In this framework, the low-temperature transition (from LS to IS) and, high-temperature transition (from IS to HS) would both be thermally activated.

Once established the general framework of the LaCoO$_3$ phenomenology, much effort has been recently devoted to obtaining *quantitative* agreement with the experiments. A typical example is the work by Saitoh and coworkers[14], aimed at fitting susceptibility, photoemission spectroscopy and x-ray absorption spectroscopy data and comparing the results with band structure calculations. In the midst of this activity, the behavior of the crystal structure has remained somewhat controversial. Asai *et al.*[8] explored the implications of the 3-level model (LS-IS-HS) for the structural properties of LaCoO$_3$, by studying the temperature evolution of the unit-cell volume of a single crystal. The fundamental *ansatz* of this analysis is that the $Co^{+3}$ ionic radius is progressively increasing with increasing spin state. Both structural and magnetic data can be well described by a varying population of LS, IS, and HS. However, Asai *et al.* found that a simple activated model is not in complete agreement with the data, and proposed a modified model containing a concentration-dependent energy contribution. The work of Asai *et al.* is undoubtedly a major step forward in understanding the structural and magnetic anomalies in $LaCoO_3$, but it also raises a number of issues. First of all, in Asai's work, the contributions of the various spin states to volume expansion are free parameters, and cannot be easily related to ionic radii. Secondly, one would expect to see anomalies not only in the unit cell volume, but also in the internal bond lengths. Furthermore, temperatures where different spin states coexist in large proportions should be characterized by significant structural disorder, which should be evident in the atomic displacement parameters (ADP). The disorder could be further enhanced by a Jahn-Teller distortion of the IS species. In fact, no such anomalies were detected in the seminal neutron work by Thornton *et al.* [10, 11]. To add further confusion to the issue, a very recent time-of-flight neutron diffraction work by Xu *et al.* [16] also indicates the *absence* of anomalies at the spin-state transitions.

The main purpose of this paper is to reconcile the current understanding of the $LaCoO_3$ spin-state transitions with state-of-the-art crystallographic data. We have refined the lattice and internal structural parameters of $LaCoO_3$ as a function of temperature, in the interval $5 \leq T \leq 1000$ K, using the Rietveld method, based on high-resolution neutron powder diffraction data. We observe clear anomalies, not only of the lattice parameters, but also of the bond lengths, associated with the two spin-state transitions. We present models of both susceptibility and thermal expansion data, based on realistic values of the ionic radii for the various $Co^{+3}$ spin states. The simplest possible scenario, based on a LS-HT transition, does not require a detail modeling of the varying HS population. Rather, one can attempt to make the susceptibility and thermal expansion data mutually consistent with a *single* adjustable parameter (the HS ionic radius, $r_{HS}$). This procedure gives a surprisingly good qualitative description of the data, and clearly indicates that a 3-level model can provide a better agreement only if $Co^{+3}_{IS}$ is significantly *smaller* than the average of $Co^{+3}_{LS}$ and $Co^{+3}_{HS}$. Two different 3-state models are presented and discussed. No ADP anomalies are detected, but, based on our estimated ionic radii and fractions of the $Co^{+3}$ spin states, we can show that these would be too small to be seen given the precision of our data. Finally, we present evidence of a *third* structural transition occurring above 800 K, which is related to the formation of oxygen vacancies.

## 2. EXPERIMENTAL

Polycrystalline $LaCoO_3$ was synthesized by solid state reaction. Starting materials of $La_2O_3$ and $Co_3O_4$ were mixed in stoichiometric proportions and heated in air at 1050° C for 5 hours, at 1100° C for 12 hours, and then 1200° C for 20 hours with intermediate grinding. Powder x-ray diffraction study shows a clean single phase pattern, and magnetization, measured with a commercial SQUID magnetometer, is consistent with the literature[8].

High-resolution neutron powder diffraction data were collected on the D2B diffractometer at the Institut Laue-Langevin (ILL), in the temperature range $5 \leq T \leq 1000$ K. As an example, the Rietveld refinement pattern for $LaCoO_3$ at 1000 K is plotted in Figure

1. Three different types of sample environment were employed. For low-temperature measurement (5≤T≤50 K), we used a helium "orange" cryostat. For the intermediate-temperature range (T≤ 500 K), we employed an ILL "cryofurnace", while for high-temperature measurements we used a standard ILL high-temperature furnace. Because the latter uses V foil heating elements, the sample is placed under high vacuum during the measurement ($P<1\times10^{-5}$ Torr). For all measurements, the sample was contained in a vanadium tube (with internal radius. 0.45 or 0.75 cm). Because of the sizable Co neutron absorption cross section (37.18 barns), an appropriate attenuation correction is crucial for obtaining correct ADPs. We applied a standard Debye-Scherrer attenuation correction to the raw data before the refinements. A residual temperature-independent systematic error of the order of $\pm$ 0.0005 Å$^2$ could still affect our refined ADPs due to the difficulty of estimating effective sample densities in a reliable way.

Rietveld refinements were carried out using the program GSAS[17]. As in the previous crystallographic study [10, 11], we could not find any evidence of the c-glide violation, which would indicate a symmetry lowering to the space group $R\bar{3}$. This violation would manifest itself with the presence of extra Bragg peaks. Therefore, all refinements were performed in the space group $R\bar{3}c$ (No. 167), using the hexagonal axes setting. As well as the x coordinate of oxygen (the only one allowed to vary in $R\bar{3}c$), we refined the oxygen site occupancy, the isotropic ADP (Debye-Waller factors) of Co and La and the anisotropic ADP tensor for oxygen. The refined structural parameters are reported in Table I. Selected bond lengths and angles are reported in Table II.

3. STRUCTURAL RESPONSE

The temperature evolution of the magnetic susceptibility and of the unit cell volume in LaCoO$_3$ has been presented in the past[8]. However, the structural response to the spin-state transition is also demonstrated by plotting some of the *internal* structural parameters. Figure 2 shows the temperature evolution of the "long" La-O bond lengthIn the  space group, the 12-fold co-ordination shell of the A-site cation is split into three groups of equal bonds: "long" (3-fold degenerate), "intermediate" (6-fold) and "short" (3-fold). In other perovskites, this parameter usually has a very weak and monotonic temperature dependence, arising from the finely balanced competition between the overall lattice expansion (which tend to expand all the bonds) and the "unilting" of the *B*O$_6$ octahedra (which tends to shorten this particular bond). This competition makes the long La-O bond particularly sensitive to non-thermal lattice distortions. The two well-known transitions at ~100 and ~500 K are both associated with rapid *increases* of the La-O "long" bond length, indicating that the structure is either becoming more distorted or it is expanding in an anomalous way (as we shall see, the latter explanation is the correct one). Both explanations are compatible with an increase of the effective ionic radius of the *B*-site, and a corresponding reduction of the Goldshmidt tolerance factor[19]. Another noteworthy feature is the rapid drop of the La-O "long" bond length above 800 K. This third transition, which is also evident in the behaviour of other parameters, such as the unit cell volume, the internal tilt angles and the oxygen ADPs (see below), is to be attributed to the formation of oxygen vacancies in the highly reducing conditions of the vacuum furnace (which is heated by vanadium metal elements). The presence of these vacancies is clearly established from the refinement of the oxygen site occupancy (Figure 3).

## 4. THERMAL EXPANSION – LATTICE CONTRIBUTION

Figure 4 shows the temperature evolution of the unit cell volume, which we express in the form of linear expansion $\alpha = \frac{1}{3}\frac{V(T)-V(0)}{V(0)}$. It is essential to obtain an accurate estimation of the three components of the lattice expansion, due to phonons, magnetism and oxygen vacancies. The phonon contribution to the thermal expansion has been evaluated using the semi-empirical formulation introduced by Ruffa[20, 21] and applied to the perovskite case by Inaba [22] (Appendix I). At high temperatures, this expression has to be corrected, due to the formation of oxygen vacancies. The thermodynamic parameters controlling the formation of oxygen vacancies in $La_{1-x}Sr_xCoO_{3-\delta}$ have been determined by Mizusaki *et al.* [23] Clearly, in our measuring conditions, the sample is not in thermodynamic equilibrium with gaseous oxygen. Nevertheless, following Mizusaki's treatment (Appendix II), we have adjusted the equilibrium oxygen partial pressure to obtain a quantitative agreement with the measured concentration of oxygen vacancies. The effect of oxygen vacancies on the unit cell volume is expressed by the coefficient $\frac{1}{V}\left(\frac{\partial V}{\partial \delta}\right)_{T,P}$, which we assume to be temperature independent. This coefficient has been refined as part of the general fitting of the volume expansion (Table III).

## 5. THERMAL EXPANSION – MAGNETIC CONTRIBUTION

In simple terms, the appearance of higher-spin states for $Co^{+3}$ contributes to the thermal expansion because these states have a larger ionic radius with respect to $Co^{+3}_{LS}$. However, the relationship between $B$ site ionic radius and unit cell volume is non-trivial, because both bond lengths and bond angles are affected. Nevertheless, it is essential to estimate, for the species in question, realistic ionic radii which, for $Co^{+3}_{LS}$ and $Co^{+3}_{HS}$, can be compared with the tabulated values, $r_{LS}$=0.545 Å and $r_{HS}$ = 0.61 Å ( Reference [24]). We therefore write the contribution of each species s to the thermal expansion as

$$\alpha_s = x_s \cdot \tfrac{1}{3}(r_s - r_{LS}) \cdot \frac{1}{V}\frac{\partial V}{\partial r} \tag{5.1}$$

where $r_s$ is the ionic radius of the species in question and $x_s$ is its fraction. We have estimated $\frac{1}{V}\frac{\partial V}{\partial r}$ from an analysis of the trend of volume versus $r$ in $LaBO_3$ ($B$= metal ion) perovskites (Appendix III), which yields a value of 0.88 Å$^{-1}$.

The magnetic contribution to the lattice expansion has been calculated in two different ways for the hypotheses of LS-HS and LS-IS-HS transitions. The purpose of the two models is quite different. In the 2-state scenario, we will make no attempt to model the HS population, which clearly has complex temperature dependence. Rather, we verify whether the susceptibility and thermal expansion data can be made to be mutually consistent by tuning $r_{HS}$ as a single parameter. For the 3-state model, we will calculate the fractions of the individual spin states following the approach used in Reference 8.

LS-HS: Assuming that both the changes in magnetic susceptibility and those of the unit cell volume are due to a varying proportion on $Co_{HS}^{+3}$, we can easily relate them to one another. In fact, the fraction of $Co_{HS}$, $x_{HS}$, can be evaluated by the formula:

$$x_{HS} = \frac{3k_B T \cdot \chi}{p_{HS}^{eff\,2}}, \qquad (5.2)$$

where $p_{HS}^{eff}$ is the effective magnetic moment of $Co_{HS}^{+3}$ (ideally 4.90 $\mu_B$). Consequently, in this model, the magnetic contribution to the thermal expansion is

$$\alpha_{mag} = x_{HS} \cdot \tfrac{1}{3}(r_{HS} - r_{LS}) \cdot \frac{1}{V}\frac{\partial V}{\partial r} \qquad (5.3)$$

It is noteworthy that no precise knowledge of the mechanism of the spin-state transition is required. We have extracted the high-spin phase fraction from published susceptibility data[8] (Fig. 5), and compared the resulting magnetic thermal expansion with our data (Fig. 6). A single-parameter fit of $r_{HS}$ was performed to optimize the agreement with the thermal expansion data, yielding a value of $r_{HS}$=0.595. This is in fairly good agreement with the Shannon value of $r_{HS}$=0.61. Clearly, the overall behavior of the calculated lattice expansion is reproduced, but there is no quantitative agreement. More importantly, the deviation from the observed lattice expansion are *opposite* to what one would expect based on the knowledge of the transport properties. In fact, we obtain a *smaller* expansion in the low-temperature *insulating* region, where the effective ionic radius of $Co_{HS}^{+3}$ is expected to be large, and a *larger* expansion in the metallic high-temperature region. This clearly indicates that, in order to obtain a better agreement with the thermal expansion data, one needs to introduce a species *other* than $Co_{HS}^{+3}$, for which the ionic radius is reduced more than the effective moment.

LS-IS-HS: To study this case, we employ Asai's[8] and Bari and Sivardière's[25] formulation (both based on a model proposed by Chestnut)[26]. We write the free energy as

$$f = PQ - \frac{1}{\beta}\ln(Z) \tag{5.4}$$

where $Z = \sum_{s=0}^{2} v_s e^{-\beta E_s}$ is the partition function and $\beta = 1/kT$. We can express the energy of each spin state as

$$E_s = \Delta_s + \frac{1}{2}\xi(Q - Q_s)^2 \tag{5.5}$$

where $Q = (V - V_0)/V_0$ and $Q_s = (V_s - V_0)/V_0$.

Here, $V_s$ is the equilibrium volume of spin state $s$ and $v_s$ is the multiplicity of each spin state, and $\xi$ is the elastic energy term is $\xi = B \cdot V_{u.c.} \approx 58$ eV. The values of $v_s$ for LS and HS are 1 and 15, respectively. Regarding the IS, multiplicity, different choices are made in the literature, depending on whether one considers the Jahn-Teller energy to be greater (no orbital degeneracy, $v_{IS} = 9$) or smaller ($e_g$ orbital degeneracy, $v_{IS} = 18$) than the relevant temperatures. Clearly, it is conceivable that the degree of orbital degeneracy may be temperature-dependent, but we have not explored this possibility. For single values of $v_{IS}$, we systematically obtain better agreement with our data in the absence of orbital degeneracy; we will therefore assume $v_{IS} = 9$ for the remainder of the paper.

The solutions for $Q$ are:

$$Q = x_1 \cdot Q_1 + x_2 \cdot Q_2 \tag{5.6}$$

where $x_s = Z^{-1} v_s e^{-\beta E'_s}$

Here, $x_1$ and $x_2$ are the concentration of the intermediate and high spin states, respectively. Similar to the treatment by Asai *et al.*, we give ourselves the option of apply correction terms to the intermediate and high-spin energy of the form:

$$E'_1 = \Delta_1 + \frac{1}{2}\xi(Q-Q_1)^2 + \Gamma_1 \cdot x_1$$
$$E'_2 = \Delta_2 + \frac{1}{2}\xi(Q-Q_2)^2 + \Gamma_2 \cdot x_2 \tag{5.7}$$

The main purpose of the correction terms is to adjust the width of the two transitions. Clearly, susceptibility data do not follow a simple activated behavior, which would predict a sharper low-temperature transition and a broader high-temperature transition in Figure 5. The presence of these terms can be rationalized in a variety of ways. Asai *et al.* argue that the energy per HS ion should decrease as their concentration increases, due to the enhanced hybridization between like species, yielding a negative value of $\Gamma_2$. Likewise, the formation of ordered phases on the local scale at low temperatures may prevent a rapid increase of the IS state concentration (*positive* $\Gamma_1$).

The magnetic susceptibility is:

$$\chi = \frac{x_{IS} \cdot p_{IS}^{eff\,2} + x_{HS} \cdot p_{HS}^{eff\,2}}{3k_B T} \tag{5.8}$$

Where the effective moments (ideally 2.83 and 4.90 $\mu_B$ for IS and HS, respectively) are also parameters of the fit. Figure 5 show the comparison of the measured magnetic susceptibility and the calculated curves for the simple activated model (dotted line) and the corrected model (solid line). The calculated lattice expansion for the corrected model is plotted in Fig. 6 (the curve for the activated model is very similar and is omitted for clarity). The fitting parameters are reported in Table III. Clearly, the fit to the lattice expansion is much better than for the 2-state model, and this is to be expected because of the greater number of adjustable parameters. For the simple activated model, refined values for the HS ionic radius (0.604 Å) is in good agreement with the Shannon value (0.61 Å). As we already remarked, the slightly reduced value for the corrected model would also be unsurprising, due to the metallic nature of the high-temperature state. For both models, the ionic radius of the IS state is closer to that of LS than to HS, as predicted based on the analysis of the simple 2-state model. The corrected model produces a distinctively better agreement with the susceptibility data, while the thermal expansion data are modeled equally well. However, some of the refined parameters, such as the greately reduced value of the effective moment ( $p_{HS}^{eff}$=3.7 $\mu_B$), seem to be somewhat unphysical. It is conceivable that even the corrected model may not fully capture the physics of the spin-state transitions.

6. BOND LENGTHS AND ANGLES

As we have seen in the previous paragraph, the spin transitions induce significant changes in the unit cell volume, which we interpret as due to an increase of the Co ionic radius. Consequently, the effect of these transitions should be directly visible in the Co-O bond lengths, while the Co-O-Co bond angles should be less affected. Ordinarily, both bond lengths and bond angles affect the unit cell volume. In the first approximation, the relationship between these parameters can be expressed as

$$\alpha = \tfrac{1}{3}\frac{\Delta V}{V_0} = \frac{\Delta d_{Co-O}}{d_{Co-O}} + \cot\frac{\psi}{2}\cdot\Delta\psi \qquad (6.1)$$

where $d_{Co-O}$ is the Co-O bond length and $\psi$ is the Co-O-Co bong angle. Equation (6.1) expresses the constraints of network connectivity in the simple hypothesis of rigid $CoO_6$ octahedra. In order to compare the relative contributions of the internal parameters, it is useful to extract the anomalous (non-phonon) part of the lattice thermal expansion from the data shown in Figure 4. The bond length contribution to the non-phonon thermal expansion is plotted in Fig. 6, together with the non-phonon (i.e., magnetic + oxygen vacancies) lattice thermal expansion. Here, we have modeled the phonon-induced thermal expansion of the Co-O bond with a single Einstein oscillator with $T_E$=700 K. From Fig. 6, it is clear that the Co-O bond expansion accounts for almost all the non-phonon lattice expansion, except at very high temperatures where oxygen vacancies are shown to play a role. Once again, this is a further confirmation that the spin-state transitions affect the structure through an increase of the Co ionic radius. For comparison, we have extracted the bond angle contribution to the non-phonon thermal expansion from Equation (6.1). Although this procedure may not be rigorously correct (strictly speaking, Equation (6.1) is only valid for the overall thermal expansion, not on the anomalous part), it clearly indicates that vacancies affect the lattice mostly through a change in the Co-O-Co bond anglesA high-temperature anomaly of the Co-O-Co bond angles is clearly visible by plotting refined bond angles from Table II .

## 7. ATOMIC DISPLACEMENT PARAMETERS

In principle, one might expect that the mixing of spin states with different ionic radii would affect the Atomic Displacement Parameters (ADPs) of cobalt and/or oxygen in a in a significant way. However, it is easy to see that, given the values of the ionic radii we obtained from the fits, these effects would be too small to be observed. For example, taking $r_{LS}=0.545$ Å $r_{LS}=0.56$ Å and $r_{HS} = 0.61$ Å, the bond length variances would be $0.6\times10^{-4}$ Å$^2$ and $0.1\times10^{-2}$ Å$^2$ for 50% mixtures of LS-IS and LS-HS, respectively. These values are too small to be observed in the relevant temperature ranges, even if they were applied entirely to one atom, rather that being distributed between Co and O. The projections of the oxygen ADP along the Co-O (which should be the most sensitive direction) and in the plane perpendicular to it, as well as the average value, are plotted in Fig. 7. The solid lines are fits up to 800 K using the Einstein model, which yield Einstein temperatures of 297 K, 500 K and 334 K, respectively. No anomaly is apparent in the temperature range $0 \leq T \leq 800$ K. On the contrary, significant deviations are apparent above 800 K, and should be attributed to the enhanced structural disorder arising from the formation of oxygen vacancies.

## 8. CONCLUSION

The observation of significant anomalies of the bond lengths in LaCoO$_3$, associated with the Co$^{+3}$ spin-state transitions, represents the central result of the present work. We have shown that the temperature dependence of the bond lengths, unit cell volume and magnetic susceptibility can be well accounted for using models based on simple energetic considerations and yielding realistic values of the ionic radii for the various spin states.

A LS-IS-HS model without IS orbital degeneracy provides the best fit to the data. However, correction terms to a simple activated behavior and a significantly reduced HS effective moment are required to obtain close agreement with the susceptibility data, indicating that the physics of the spin-state transitions may not be completely captured by these models. We have shown that the ADP anomalies, which would be expected in the presence of coexisting species with different ionic radii, are too small to be detected with the current precision. We have also observed a *third* structural anomaly at higher temperatures, which we have explained with the formation of oxygen vacancies.

## APPENDIX I: PHONON CONTRIBUTION TO THE THERMAL EXPANSION

The phonon contribution to the thermal expansion has been evaluated using the semi-empirical formulation introduced by Ruffa[20, 21] and applied to the perovskite case by Inaba[22], as the sum of two terms

$$\left(\frac{\Delta l}{l}\right) = \left(\frac{\Delta l}{l}\right)_1 + \left(\frac{\Delta l}{l}\right)_2 \tag{AI.1}$$

where

$$\left(\frac{\Delta l}{l}\right)_1 = 3 \cdot \left(\frac{1}{2ar_n D}\right) \cdot (k_B T) \cdot \left(\frac{T}{\Theta_D}\right)^3 \int_0^{\frac{\Theta_D}{T}} \frac{x^3 dx}{e^x - 1} \tag{AI.2}$$

(AI.2) has the same expression as the Debye-Grüneisen thermal expansion with the substitution $2ar_n D = \frac{BV_m}{\gamma n}$, whereas

$$\left(\frac{\Delta l}{l}\right)_2 = \frac{3}{8} \cdot \left(\frac{1}{ar_n D^2}\right) \cdot (k_B T)^2 \cdot \left(\frac{T}{\Theta_D}\right)^3 \int_0^{\frac{\Theta_D}{T}} \frac{x^4 (1+e^x) dx}{(e^x - 1)^2} \qquad \text{(AI.4)}$$

Here $\gamma$ is the Grüneisen parameter (~2 for perovskites), $B$ is the bulk modulus (typically 150-180×10$^9$ nm$^{-2}$ for perovskites) and $V_m$ is the molar volume (3.3×10$^{-5}$ m$^3$mol$^{-1}$ for LaCoO$_3$), $k_B$ is the Boltzmann's constant ( = 8.31 Joules·mol$^{-1}$·K$^{-1}$), $\theta_D$ is the Debye temperature and $n$ is the number of atoms per formula unit (5 in our case). The parameters of Ruffa's formulation are the depth and inverse width of the Morse potential $D$ and $a$ and the mean atomic separation $r_n$. We have used the values determined by Inaba et al. [22] for LaMnO$_3$ ($a$ = 2.341 Å, $r_n$ = 0.721 Å$^{-1}$, $D$ = 211.9 kJmol$^{-1}$ and $\theta_D$=620.2 K) but we have rescaled the product $a \cdot r$ so that the high-temperature value of the thermal expansion coefficient matches the experimental value for La$_{0.92}$Sr$_{0.08}$CoO$_3$[8].

## APPENDIX II: OXYGEN VACANCIES CONTRIBUTION TO THE THERMAL EXPANSION

The equilibrium condition of the crystal lattice with gaseous oxygen is expressed by the equation:

$$\mu - \mu_0 = h - h_0 - T(s - s_0) = \frac{RT}{2} \ln PO_2 \qquad \text{(AII.1)}$$

where $R$ = 8.314 JK$^{-1}$mol$^{-1}$ is the gas constant and $PO_2$ is the oxygen partial pressure (in Atm). Misuzaki et al. [23] have determined the following expressions for the enthalpy and entropy of oxygen vacancy formation:

$$h - h_0 = \Delta h_0^0 - a\delta \quad \text{and} \quad s - s_0 = \Delta s_0^0 + R\ln\left(\frac{\delta}{3-\delta}\right) \quad \text{(AII.2)}$$

where $\delta$ is the vacancy content per unit formula. For $LaCoO_{3-\delta}$, $\Delta h_0^0 =$ -2095 KJmol$^{-1}$, $\Delta s_0^0 =$ -105 Jmol$^{-1}$ and a=2092 KJmol$^{-1}$. The oxygen vacancy content at a given $PO_2$ can be determined by solving the following equation with respect to $\delta$:

$$\frac{e^{\frac{\Delta h_0^0 - a\delta}{RT} - \frac{\Delta s_0^0}{R}}}{\frac{\delta}{3-\delta}} - (PO_2)^{1/2} = 0 \quad \text{(AII.3)}$$

It can be easily seen that with the approximation $a \approx 0$ and $3-\delta \approx 3$ one recovers a simple Arrhenius law. Data in Figure 3 have been fitted by performing a numerical minimization of Equation (AII.3) as a function of $PO_2$.

## APPENDIX III: VOLUME VS. IONIC RADIUS IN LANTHANUM OXIDE PEROVSKITES

We have estimated $\frac{1}{V}\frac{\partial V}{\partial r}$ from an analysis of the trend of volume versus $r$ in La$BO_3$ perovskites, where $B$ is a transition metal, Al or Ga. The data were obtained by mining the Inorganic Crystal Structure Database[28, 29] (Table IV). For small $B$ site cations, the structure is rhombohedral (space group $R\bar{3}c$), and a transition to the orthorhombic space group $Pnma$ occurs for $r_B \sim 0.6$. LaCuO$_3$ is clearly anomalous (the properties of this high-pressure compound are somewhat dubious), and was excluded from the fit, which yielded a value $\frac{1}{V}\frac{\partial V}{\partial r} = 0.88$.

## Figure Captions

Figure 1: An example of Rietveld neutron powder diffraction pattern for LaCoO$_3$. The present data were collected at 1000 K using a standard ILL furnace. Crosses (+) represent experimental points, while the fit is shown as a solid line. A difference curve (observed minus calculated) is plotted at the bottom. Vertical tick marks indicate the position of allowed Bragg peaks in the $R\bar{3}c$ space group.

Figure 2: Temperature dependence of the "long" La-O bond length in LaCoO$_3$, as determined from neutron powder diffraction data. The solid line is a guide to the eye.

Figure 3: Oxygen vacancy concentration per formula unit δ, as determined from neutron powder diffraction data. The solid line is a fit to the data, following the treatment of Misuzaki et al. [23]. The zero-temperature δ, 0.06(2), was subtracted from the data before the fitting.

Figure 4: **Circles**: linear thermal expansion α of LaCoO$_3$, obtained from unit-cell volumes as $\alpha = \frac{1}{3}\frac{V(T)-V(0)}{V(0)}$. The solid line is a fit to the data using an activated 3-level model (LS,IS and HS) with correction terms (**dashed line**), as explained in the text. The phonon contribution (**dotted line**) was evaluated using the approach of Inaba [22], and contains no fitting parameters. The oxygen-vacancy contribution (**dash-dotted line**) was calculated by fitting the oxygen-vacancy data of Figure 3. The only additional adjustable

parameter is dlogV/dδ, which expresses the dependence of the unit cell volume on the vacancy concentration (see also Table III).

Figure 5: Fit to the magnetic susceptibility data of Reference 8 (**circles**) using the simple 3-state activated model 1(**dashed line**) and the corrected model 2 (**solid line**). The low-temperature paramagnetic behavior was subtracted as in Reference 8.

Figure 6: Spin-state anomalies of the LaCoO$_3$ unit cell and internal parameters. **Circles**: the non-phonon component of the Co-O bond length expansion, obtained from the Rietveld bond length values by subtracting a single Einstein oscillator with $T_E$=700 K. **Diamonds**: the non-phonon component of the linear (unit-cell) thermal expansion, obtained as in Figure 4. The high-temperature deviation, due to the formation of oxygen vacancies, is clearly visible. The **solid** and **dashed** lines are the pure magnetic thermal expansion and the magnetic+vacancy thermal expansion, respectively. The Co-O-Co bond angle contribution (**squares**) was obtained from Equation (6.1).

Figure 7: Oxygen Displacement Parameters in LaCoO$_3$. Average value (**circles**) and projections of the oxygen displacement parameters *parallel* (**squares**) and *perpendicular* (**triangles**) to the Co-O bond. The solid lines are fit to the data up to 800 K, using a simple Einstein oscillator. The Einstein temperatures were 297 K, 500 K and 334 K, respectively. Clearly visible is a high-temperature upturn, probably due to additional disorder from oxygen vacancies.

Table I: Structural parameters of LaCoO$_3$ as a function of temperature, as refined from high-resolution neutron powder diffraction data. The space group is $R\bar{3}c$ (No. 167). The atomic positions are: La 6a(0,0,1/4), Co 6b(0,0,0), O 18e(x,0,1/4). Numbers in parentheses are statistical errors of the last significant digit. The last row is the fractional occupancy of the oxygen site, and is related to the vacancy concentration by the formula $\delta = 3 \cdot (1 - frac)$.

| Temperature(K) | | 5 | 10 | 20 | 30 | 50 | 100 | 125 | 150 |
|---|---|---|---|---|---|---|---|---|---|
| a (Å) | | 5.42625(5) | 5.42613(5) | 5.42617(5) | 5.42583(7) | 5.42784(6) | 5.43317(5) | 5.43530(5) | 5.43733(5) |
| c (Å) | | 12.991(1) | 12.991(1) | 12.991(1) | 12.991(2) | 12.999(1) | 13.022(1) | 13.033(1) | 13.044(1) |
| La | U 100(Å$^2$) | 0.00(2) | 0.03(2) | 0.02(2) | -0.01(3) | 0.02(2) | 0.04(2) | 0.16(2) | 0.19(2) |
| Co | U×100(Å$^2$) | 0.10(4) | 0.07(2) | 0.11(4) | 0.03(5) | 0.16(4) | 0.22(4) | 0.42(4) | 0.37(4) |
| O | x | 0.55265(8) | 0.55255(8) | 0.55256(8) | 0.55243(8) | 0.55277(8) | 0.55260(8) | 0.55216(8) | 0.55200(8) |
| | U$_{11}$×100(Å$^2$) | 0.09(2) | 0.07(2) | 0.07(2) | 0.06(2) | 0.17(2) | 0.13(2) | 0.21(2) | 0.25(2) |
| | U$_{22}$×100(Å$^2$) | 0.12(3) | 0.16(3) | 0.13(3) | 0.15(3) | 0.13(3) | 0.11(3) | 0.20(3) | 0.26(3) |
| | U$_{33}$×100(Å$^2$) | 0.33(3)00 | 0.43(3) | 0.43(3) | 0.53(3) | 0.22(2) | 0.08(3) | 0.19(2) | 0.26(2) |
| | U$_{12}$×100(Å$^2$) | 0.06(1) | 0.08(1) | 0.07(1) | 0.08(2) | 0.06(1) | 0.05(2) | 0.10(1) | 0.13(1) |
| | U$_{13}$×100(Å$^2$) | -0.06(1) | -0.04(1) | -0.04(1) | -0.05(1) | -0.09(1) | -0.11(1) | -0.13(1) | -0.14(1) |
| | U$_{23}$×100(Å$^2$) | -0.12(2) | -0.09(2) | -0.09(2) | -0.10(3) | -0.18(3) | -0.21(6) | -0.27(2) | -0.28(2) |
| | frac. | 0.983(4) | 0.985(4) | 0.986(4) | 0.994(5) | 0.982(5) | 0.971(5) | 0.965(4) | 0.968(4) |

| Temperature(K) | | 200 | 250 | 300 | 350 | 400 | 450 | 500 | 550 |
|---|---|---|---|---|---|---|---|---|---|
| a (Å) | | 5.44108(5) | 5.44490(4) | 5.44864(3) | 5.45284(4) | 5.45702(4) | 5.46230(6) | 5.46815(5) | 5.47499(6) |
| c (Å) | | 13.064(1) | 13.084(1) | 13.1035(9) | 13.123(1) | 13.142(1) | 13.163(1) | 13.185(1) | 13.207(1) |
| La | U×100(Å$^2$) | 0.19(2) | 0.29(2) | 0.36(2) | 0.46(2) | 0.50(2) | 0.62(2) | 0.69(2) | 0.84(2) |
| Co | U×100(Å$^2$) | 0.26(4) | 0.34(4) | 0.37(4) | 0.49(4) | 0.47(4) | 0.67(4) | 0.56(4) | 0.59(4) |
| O | x | 0.55134(8) | 0.55095(8) | 0.55032(8) | 0.55011(7) | 0.54956(7) | 0.54940(9) | 0.54901(8) | 0.54909(9) |
| | U$_{11}$×100(Å$^2$) | 0.31(2) | 0.44(2) | 0.51(2) | 0.65(2) | 0.73(2) | 0.88(2) | 0.98(2) | 1.15(2) |
| | U$_{22}$×100(Å$^2$) | 0.34(3) | 0.38(3) | 0.51(3) | 0.56(3) | 0.58(3) | 0.80(3) | 0.87(3) | 0.97(3) |
| | U$_{33}$×100(Å$^2$) | 0.39(2) | 0.53(2) | 0.76(2) | 0.87(2) | 0.95(2) | 1.17(3) | 1.33(3) | 1.61(3) |
| | U$_{12}$×100(Å$^2$) | 0.17(1) | 0.19(1) | 0.25(1) | 0.28(1) | 0.29(1) | 0.40(2) | 0.44(2) | 0.49(2) |
| | U$_{13}$×100(Å$^2$) | -0.14(1) | -0.17(1) | -0.24(1) | -0.27(1) | -0.29(1) | -0.34(1) | -0.34(1) | -0.36(1) |
| | U$_{23}$×100(Å$^2$) | -0.27(2) | -0.34(2) | -0.48(2) | -0.55(2) | -0.57(2) | -0.69(3) | -0.68(2) | -0.72(3) |
| | frac. | 0.974(4) | 0.976(4) | 0.975(4) | 0.980(4) | 0.979(4) | 0.975(4) | 0.978(4) | 0.971(4) |

| Temperature(K) | | 600 | 650 | 700 | 750 | 800 | 850 | 900 | 950 | 1000 |
|---|---|---|---|---|---|---|---|---|---|---|
| a (Å) | | 5.48118(6) | 5.48677(6) | 5.49250(6) | 5.49739(6) | 5.50253(7) | 5.50759(7) | 5.51221(8) | 5.51717(9) | 5.5230(1) |
| c (Å) | | 13.229(2) | 13.250(2) | 13.272(2) | 13.292(2) | 13.314(2) | 13.336(2) | 13.356(2) | 13.380(2) | 13.408(3) |
| La | U×100(Å$^2$) | 0.88(2) | 1.04(2) | 1.12(2) | 1.24(2) | 1.38(3) | 1.60(3) | 1.80(3) | 1.93(3) | 2.14(3) |
| Co | U×100(Å$^2$) | 0.78(4) | 0.92(5) | 1.04(5) | 0.87(4) | 0.97(5) | 1.01(5) | 1.26(5) | 1.42(5) | 1.44(6) |
| O | x | 0.54849(9) | 0.54804(9) | 0.54787(9) | 0.5474(1) | 0.5469(1) | 0.5466(1) | 0.5459(1) | 0.5449(1) | 0.5434(1) |
| | U$_{11}$×100(Å$^2$) | 1.24(2) | 1.46(2) | 1.59(2) | 1.83(2) | 2.05(2) | 2.27(3) | 2.51(3) | 2.67(3) | 2.92(4) |
| | U$_{22}$×100(Å$^2$) | 1.08(3) | 1.29(4) | 1.40(4) | 1.51(4) | 1.65(4) | 1.76(4) | 1.93(5) | 2.10(5) | 2.18(6) |
| | U$_{33}$×100(Å$^2$) | 1.76(3) | 2.00(3) | 2.18(3) | 2.24(3) | 2.50(3) | 2.61(4) | 3.05(4) | 3.47(4) | 4.10(5) |
| | U$_{12}$×100(Å$^2$) | 0.54(2) | 0.64(2) | 0.70(2) | 0.76(2) | 0.82(2) | 0.88(2) | 0.96(2) | 1.05(2) | 1.09(3) |
| | U$_{13}$×100(Å$^2$) | -0.43(1) | -0.45(1) | -0.52(1) | -0.56(2) | -0.57(2) | -0.62(2) | -0.64(2) | -0.74(2) | -0.82(3) |
| | U$_{23}$×100(Å$^2$) | -0.86(3) | -0.91(3) | -1.04(3) | -1.12(3) | -1.13(3) | -1.24(3) | -1.29(4) | -1.47(4) | -1.64(5) |
| | frac. | 0.971(4) | 0.971(4) | 0.967(4) | 0.967(4) | 0.973(4) | 0.961(4) | 0.958(4) | 0.955(4) | 0.948(4) |

24Table II: Selected bond lengths (in Ångstroms) and bond angles (in degrees) for LaCoO$_3$ as a function of temperature, as refined from high-resolution neutron powder diffraction data. Numbers after the "×" sign are bond multiplicities. Numbers in parentheses are statistical errors of the last significant digit.

| Temperature (K) | La-O × 3 | La-O × 3 | La-O × 6 | Co-O × 6 | O-Co-O | Co-O-Co |
|---|---|---|---|---|---|---|
| 5 | 2.42755(5) | 2.9988(5) | 2.68767(5) | 1.92544(7) | 91.477(2) | 162.93(3) |
| 10 | 2.42779(4) | 2.9982(4) | 2.68755(5) | 1.92533(7) | 91.475(2) | 162.96(3) |
| 20 | 2.42779(4) | 2.9983(4) | 2.68759(5) | 1.92535(7) | 91.475(2) | 162.96(3) |
| 30 | 2.42784(4) | 2.9974(5) | 2.68740(6) | 1.92515(8) | 91.471(2) | 163.00(3) |
| 50 | 2.42775(5) | 3.0003(4) | 2.68898(5) | 1.92627(7) | 91.470(2) | 162.90(3) |
| 100 | 2.42308(5) | 3.0024(5) | 2.69300(5) | 1.92854(7) | 91.435(2) | 162.96(3) |
| 125 | 2.43341(4) | 3.0012(4) | 2.69458(5) | 1.92922(6) | 91.409(2) | 163.10(3) |
| 150 | 2.43359(4) | 3.0014(4) | 2.69626(5) | 1.93007(6) | 91.389(2) | 163.15(3) |
| 200 | 2.44124(4) | 2.9999(4) | 2.69925(5) | 1.93141(6) | 91.342(2) | 163.37(3) |
| 250 | 2.44450(4) | 2.9999(4) | 2.70240(5) | 1.93300(6) | 91.302(2) | 163.50(2) |
| 300 | 2.45024(4) | 2.9985(4) | 2.70524(4) | 1.93452(6) | 91.260(2) | 163.71(2) |
| 350 | 2.45324(4) | 2.9996(4) | 2.70854(4) | 1.93607(6) | 91.227(2) | 163.77(2) |
| 400 | 2.45814(4) | 2.9989(4) | 2.71154(4) | 1.93757(6) | 91.189(2) | 163.95(2) |
| 450 | 2.46155(5) | 3.0010(5) | 2.71522(5) | 1.93972(7) | 91.163(2) | 164.01(3) |
| 500 | 2.46604(4) | 3.0021(4) | 2.71888(4) | 1.94183(6) | 91.134(2) | 164.13(2) |
| 550 | 2.46877(5) | 3.0063(5) | 2.72316(5) | 1.94462(7) | 91.118(2) | 164.11(3) |
| 600 | 2.47485(5) | 3.0064(5) | 2.72689(5) | 1.94671(7) | 91.084(2) | 164.30(3) |
| 650 | 2.47985(5) | 3.0070(5) | 2.73040(6) | 1.94869(7) | 91.054(2) | 164.45(3) |
| 700 | 2.48335(5) | 3.0092(5) | 2.73422(6) | 1.95098(7) | 91.029(2) | 164.51(3) |
| 750 | 2.48815(5) | 3.0093(5) | 2.73759(6) | 1.95279(7) | 90.995(2) | 164.66(3) |
| 800 | 2.49315(5) | 3.0095(6) | 2.74108(6) | 1.95468(7) | 90.960(2) | 164.82(3) |
| 850 | 2.49726(6) | 3.0104(6) | 2.74476(6) | 1.95669(8) | 90.925(2) | 164.93(3) |
| 900 | 2.50326(6) | 3.0091(6) | 2.74799(6) | 1.95828(8) | 90.884(2) | 165.16(4) |
| 950 | 2.51077(7) | 3.0064(7) | 2.75153(7) | 1.95991(9) | 90.832(3) | 165.47(4) |
| 1000 | 2.52198(8) | 3.0011(8) | 2.75549(8) | 1.9615(1) | 90.764(3) | 165.97(5) |

Table III: fitted parameters for the LS-IS-HS models. Model 1 is a simple 3-state activated behavior. Model 2 includes the correction terms $\Gamma_1$ and $\Gamma_2$, as described in the text. The parameter $d\log V/d\delta$ expresses the dependence of the unit cell volume on the vacancy concentration, and is needed to fit the thermal expansion data at high temperatures (see text). All the parameters, excluding the effective moments, were first fitted against the thermal expansion data. The effective moments were subsequently fitted against the susceptibility data (the $\chi^2$ values only refers to the latter fit).

|  | E1 (meV) | E2 (meV) | $r_{IS}$ (Å) | $r_{IH}$ (Å) | $d\log V/d\delta$ | $\Gamma_1$ | $\Gamma_2$ | $p_{IS}$ | $p_{HS}$ | $\chi^2$ |
|---|---|---|---|---|---|---|---|---|---|---|
| Model 1 | 22.41(1) | 85.45(1) | 0.5582(2) | 0.6041(6) | 0.017(1) | 0 | 0 | 2.75 | 4.53 | 107 |
| Model 2 | 13.050(3) | 136.118(3) | 0.56087(4) | 0.5824(1) | 0.059(1) | 30.351(4) | 165.435(4) | 2.95 | 3.72 | 20 |

26TABLE IV: Reference table for the La$B$O$_3$ compounds used for the determination of $\frac{1}{V}\frac{\partial V}{\partial r}$. The second column refers to the Inorganic Crystal Structure Database entry number[28, 29]. The $M$-site ionic radii are from Shannon[24].

| Compound | ICSD Entry No. | I.R. | Volume/f.u.(Å$^3$) | Symmetry |
|---|---|---|---|---|
| LaCuO$_3$ | 73554 | 0.54 | 57.967 | $R\bar{3}c$ |
| LaNiO$_3$ | 84933 | 0.56 | 56.567 | |
| LaCoO$_3$ | 201763 | 0.545 | 56.000 | |
| LaAlO$_3$ | 74494 | 0.535 | 54.467 | |
| LaCrO$_3$ | 79344 | 0.615 | 58.591 | Pnma |
| LaTiO$_3$ | 63575 | 0.67 | 62.426 | |
| LaMnO$_3$ | 50334 | 0.6645 | 61.387 | |
| LaFeO$_3$ | 84941 | 0.645 | 60.715 | |
| LaGaO$_3$ | 73760 | 0.62 | 59.043 | |
| LaVO$_3$ | 73898 | 0.64 | 60.525 | |
| LaRuO$_3$ | 75569 | 0.68 | 62.200 | |
| LaRhO$_3$ | 33122 | 0.665 | 61.957 | |

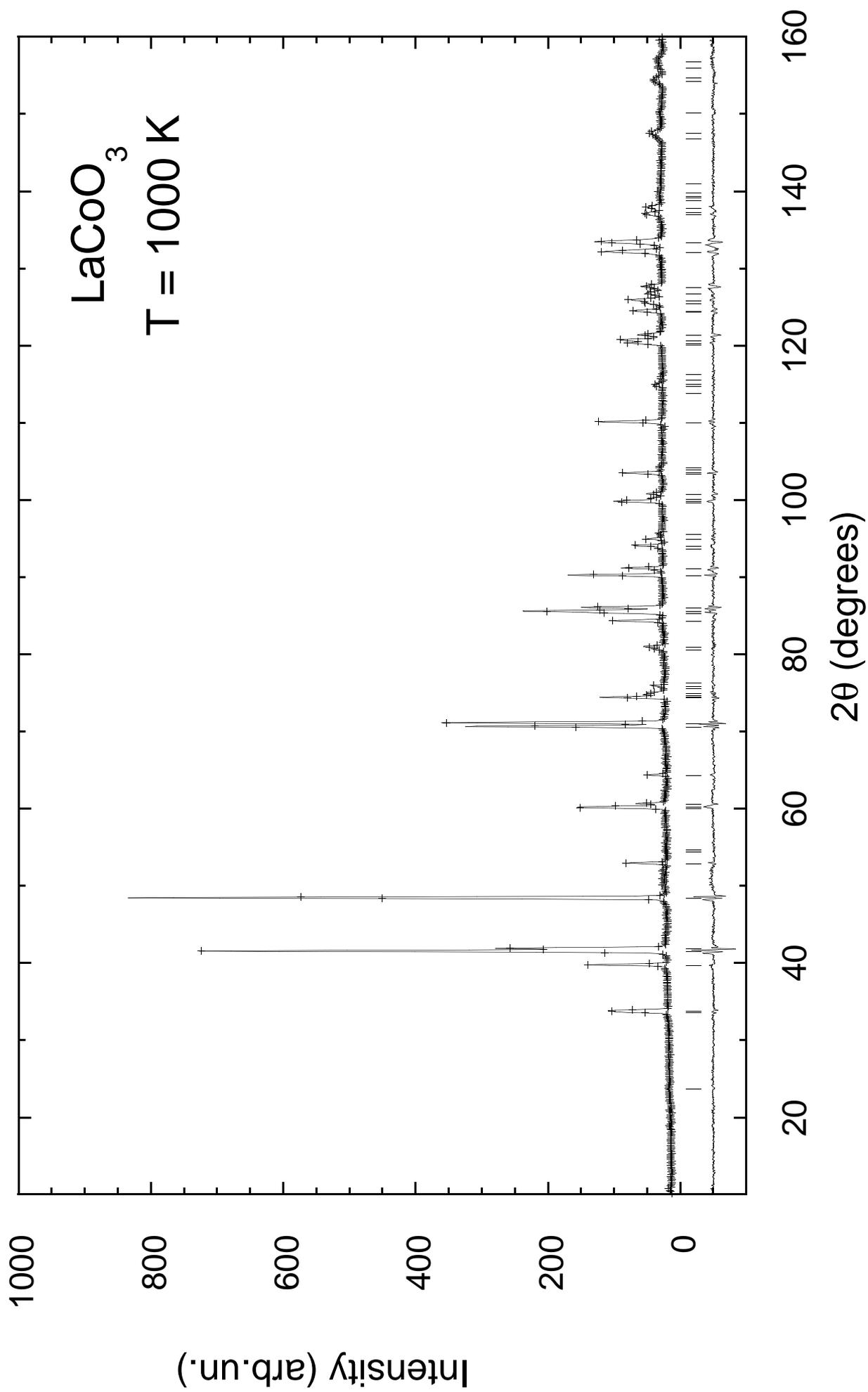

Figure 1
P.G. Radaelli & S-W. Cheong

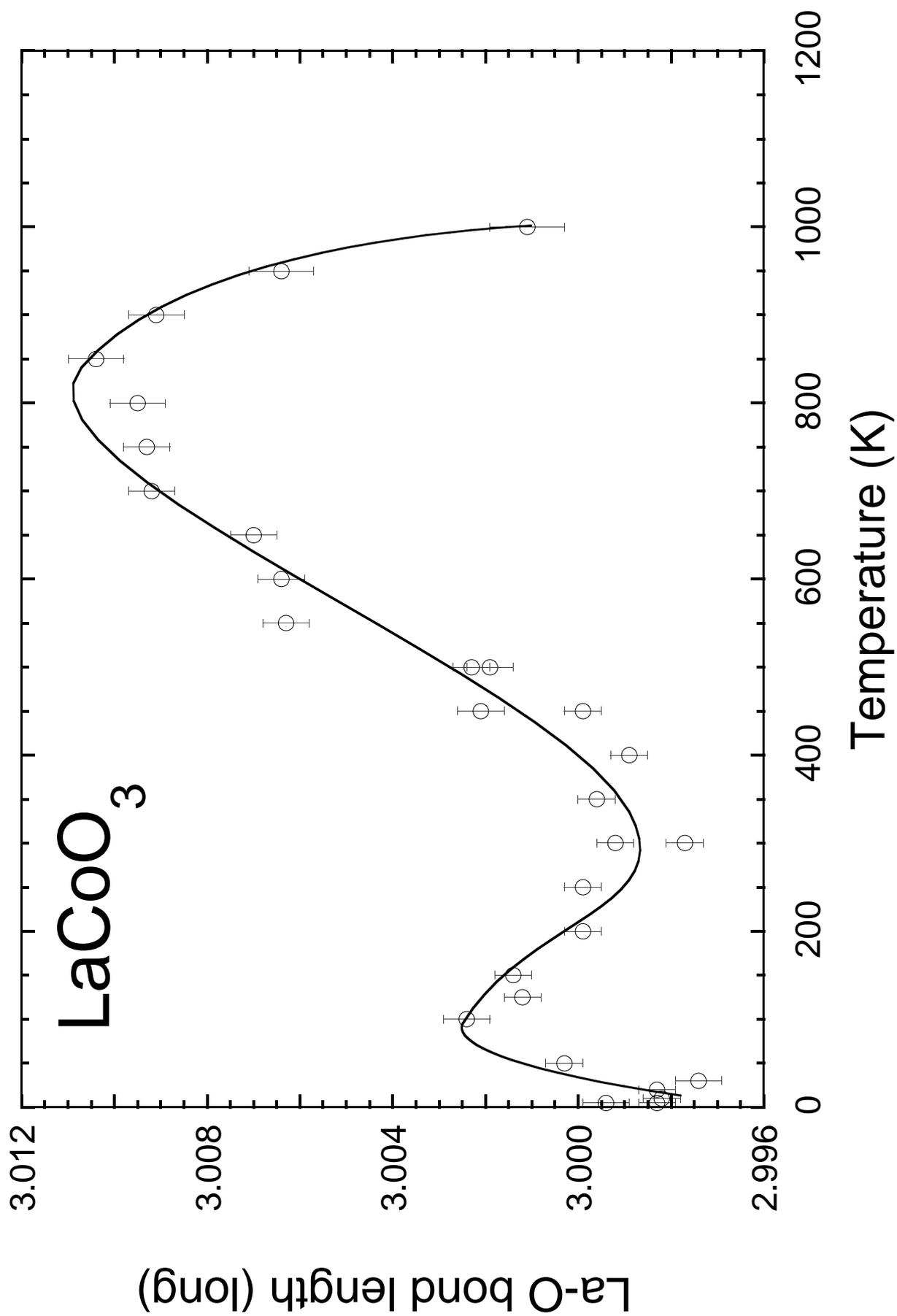

Figure 2
P.G. Radaelli & S-W. Cheong

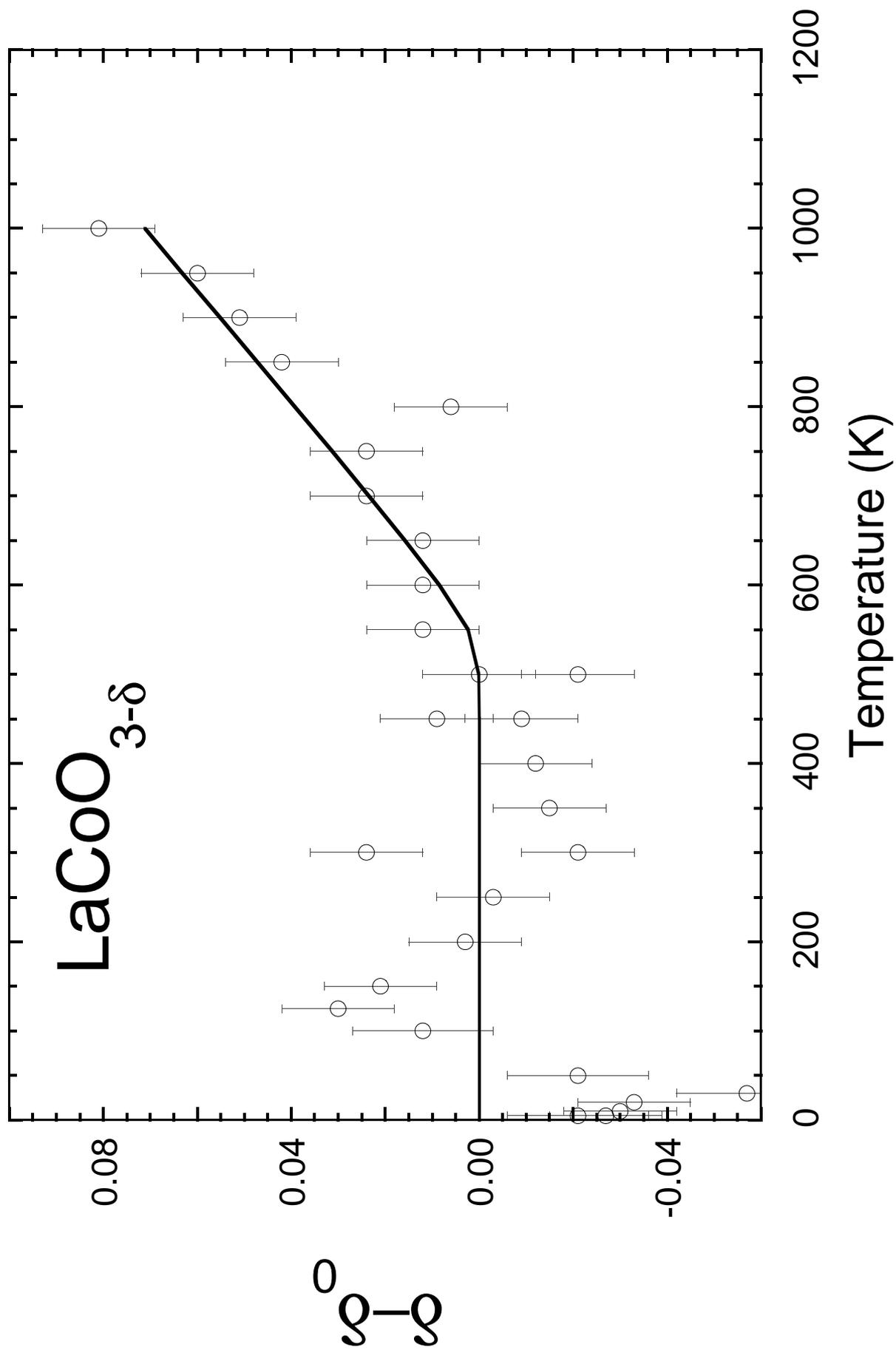

Figure 3
P.G. Radaelli & S-W. Cheong

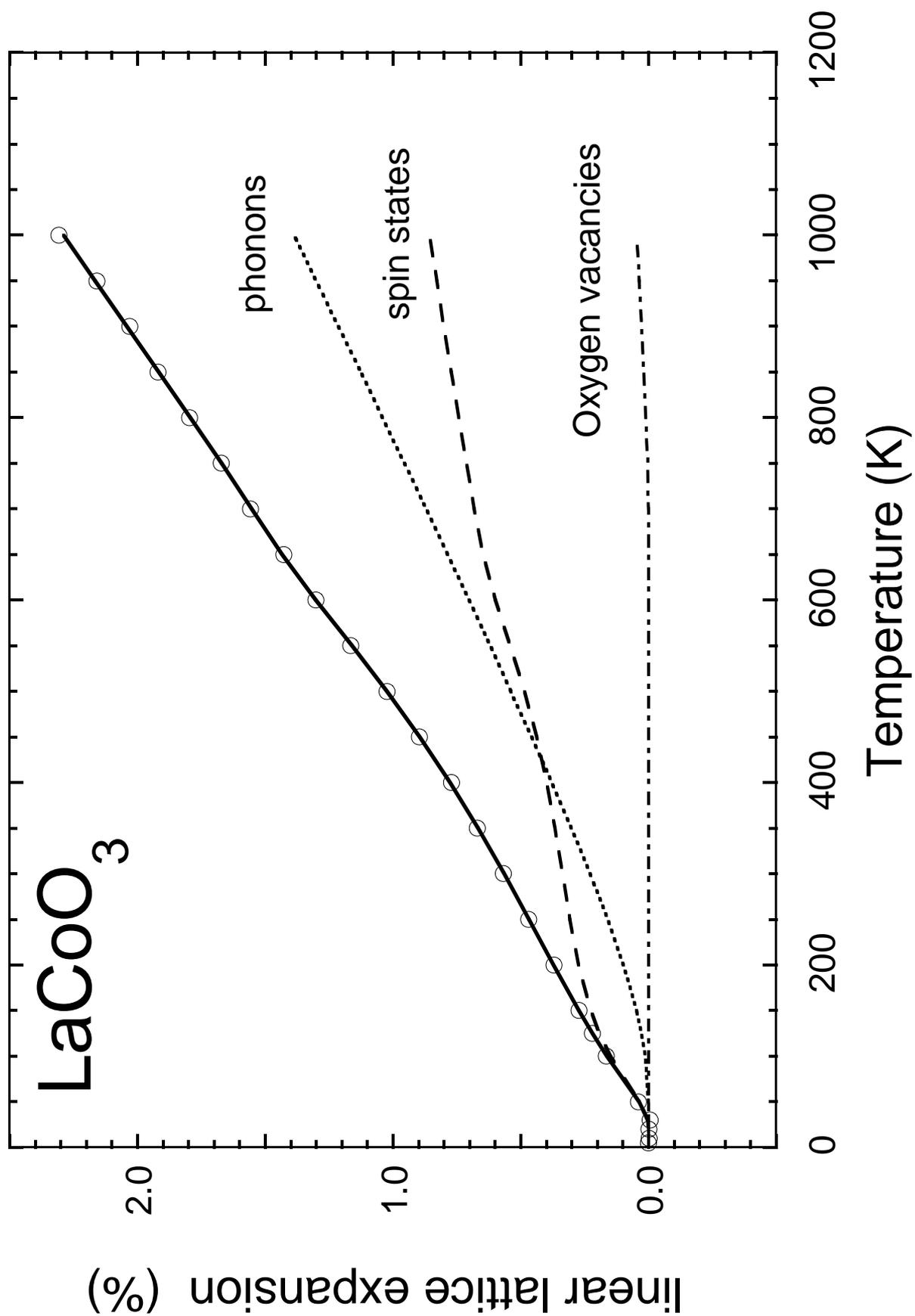

Figure 4
P.G. Radaelli & S-W. Cheong

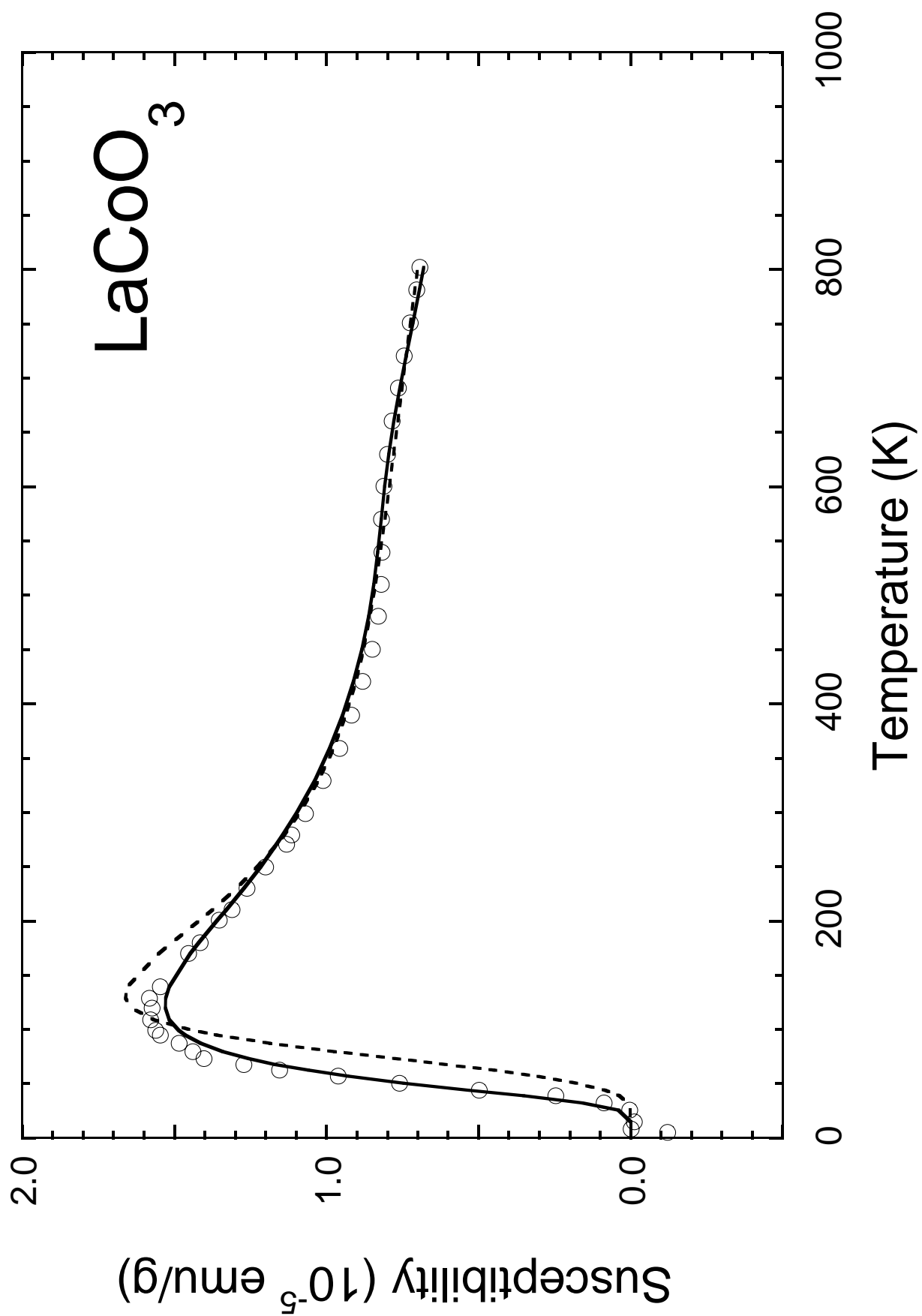

Figure 5
P.G. Radaelli & S-W. Cheong

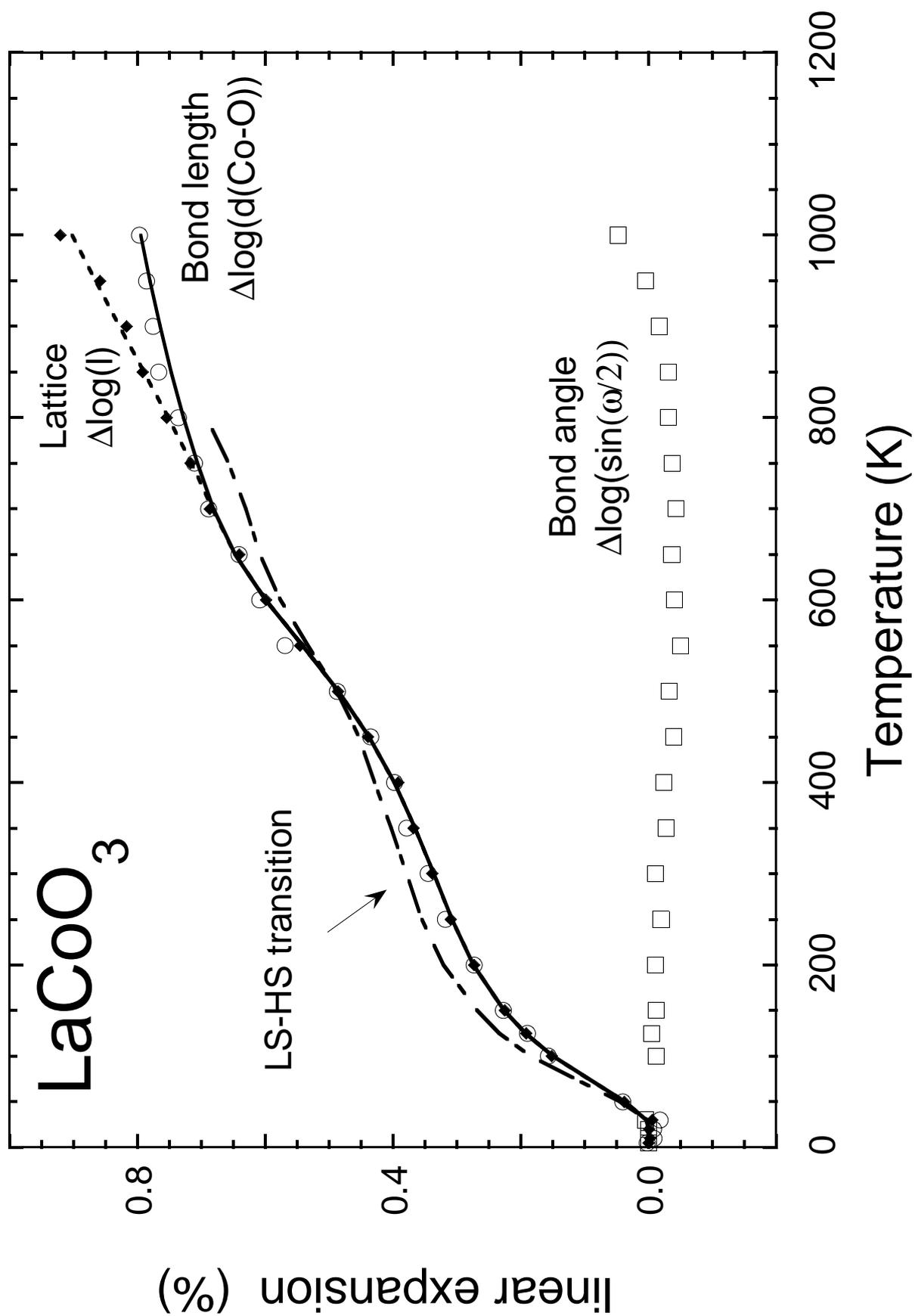

Figure 6
P.G. Radaelli & S-W. Cheong

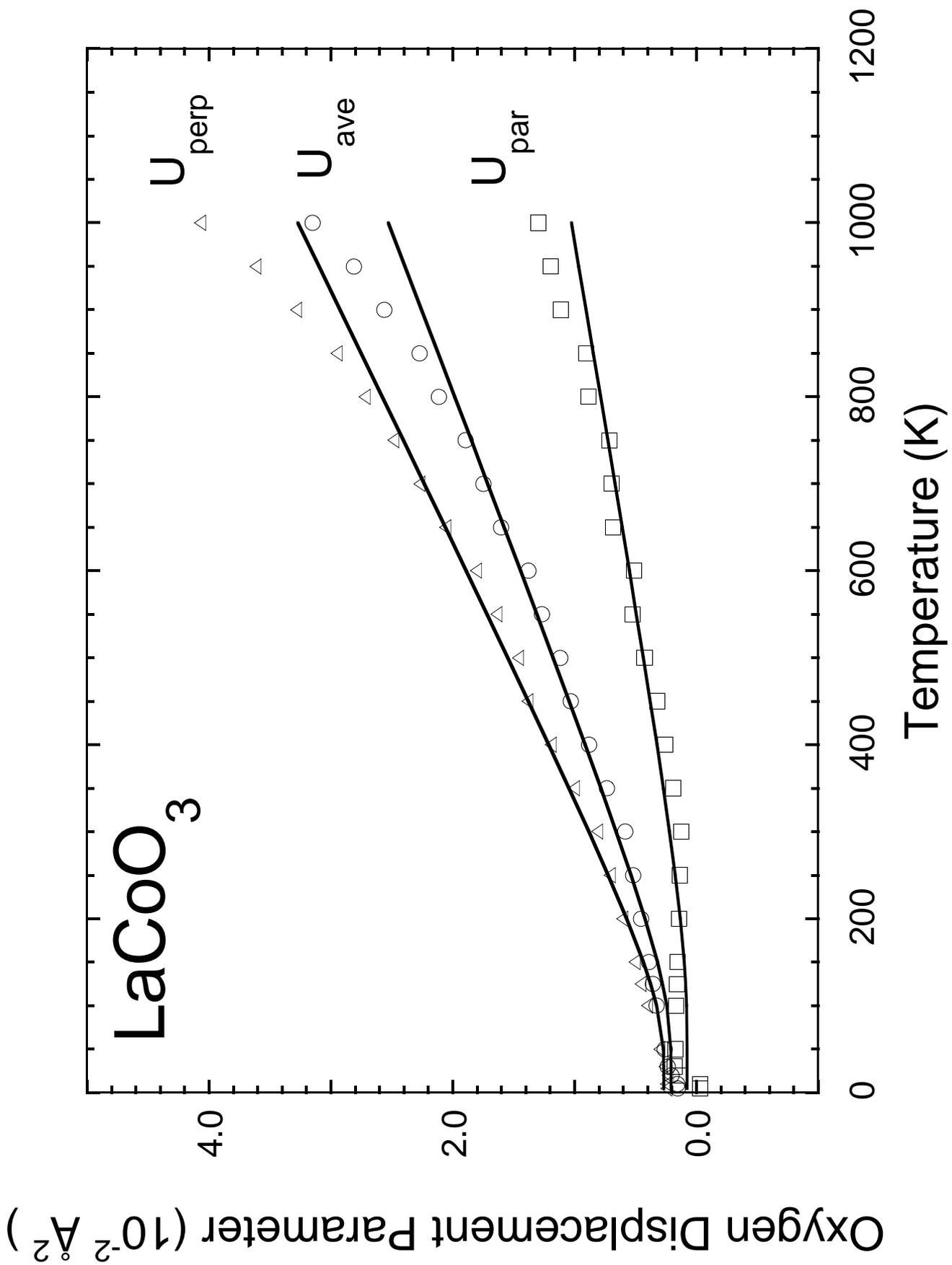

Figure 7
P.G. Radaelli & S-W. Cheong